\documentclass[manuscript]{acmart}
\AtBeginDocument{%
  }

\setcopyright{acmlicensed}
\copyrightyear{2026}
\acmYear{2026}
\acmDOI{XXXXXXX.XXXXXXX}
\acmConference[CHI EA '26]{Extended Abstracts of the CHI Conference on Human Factors in Computing Systems}{April 13--17, 2026}{Barcelona, Spain}
\acmISBN{978-1-4503-XXXX-X/2026/06}

\usepackage{subcaption}
\newcommand{\ssection}[1]{\noindent {\bf #1} }

\begin{document}

\title{Reality Copilot: Voice-First Human–AI Collaboration in Mixed Reality Using Large Multimodal Models}

\author{Liuchuan Yu}
\email{lyu20@gmu.edu}
\orcid{0000-0003-2375-1862}
\affiliation{%
  \institution{George Mason University}
  \city{Fairfax}
  \state{Virginia}
  \country{USA}
}

\author{Yongqi Zhang}
\affiliation{%
  \institution{George Mason University}
  \city{Fairfax}
  \state{Virginia}
  \country{USA}
}
\email{yzhang59@gmu.edu}

\author{Lap-Fai Yu}
\affiliation{%
  \institution{George Mason University}
  \city{Fairfax}
  \state{Virginia}
  \country{USA}
}
\email{craigyu@gmu.edu}

\renewcommand{\shortauthors}{Yu et al.}

\begin{abstract}

Large Multimodal Models (LMMs) have shown strong potential for assisting users in tasks, such as programming, content creation, and information access, yet their interaction remains largely limited to traditional interfaces such as desktops and smartphones. Meanwhile, advances in mixed reality (MR) hardware have enabled applications that extend beyond entertainment and into everyday use. However, most existing MR systems rely primarily on manual input (e.g., hand gestures or controllers) and provide limited intelligent assistance due to the lack of integration with large-scale AI models. We present Reality Copilot, a voice-first human–AI assistant for mixed reality that leverages LMMs to enable natural speech-based interaction. The system supports contextual understanding of physical environments, realistic 3D content generation, and real-time information retrieval. In addition to in-headset interaction, Reality Copilot facilitates cross-platform workflows by generating context-aware textual content and exporting generated assets. This work explores the design space of LMM-powered human–AI collaboration in mixed reality.
\end{abstract}


\begin{teaserfigure}
  \includegraphics[width=\textwidth]{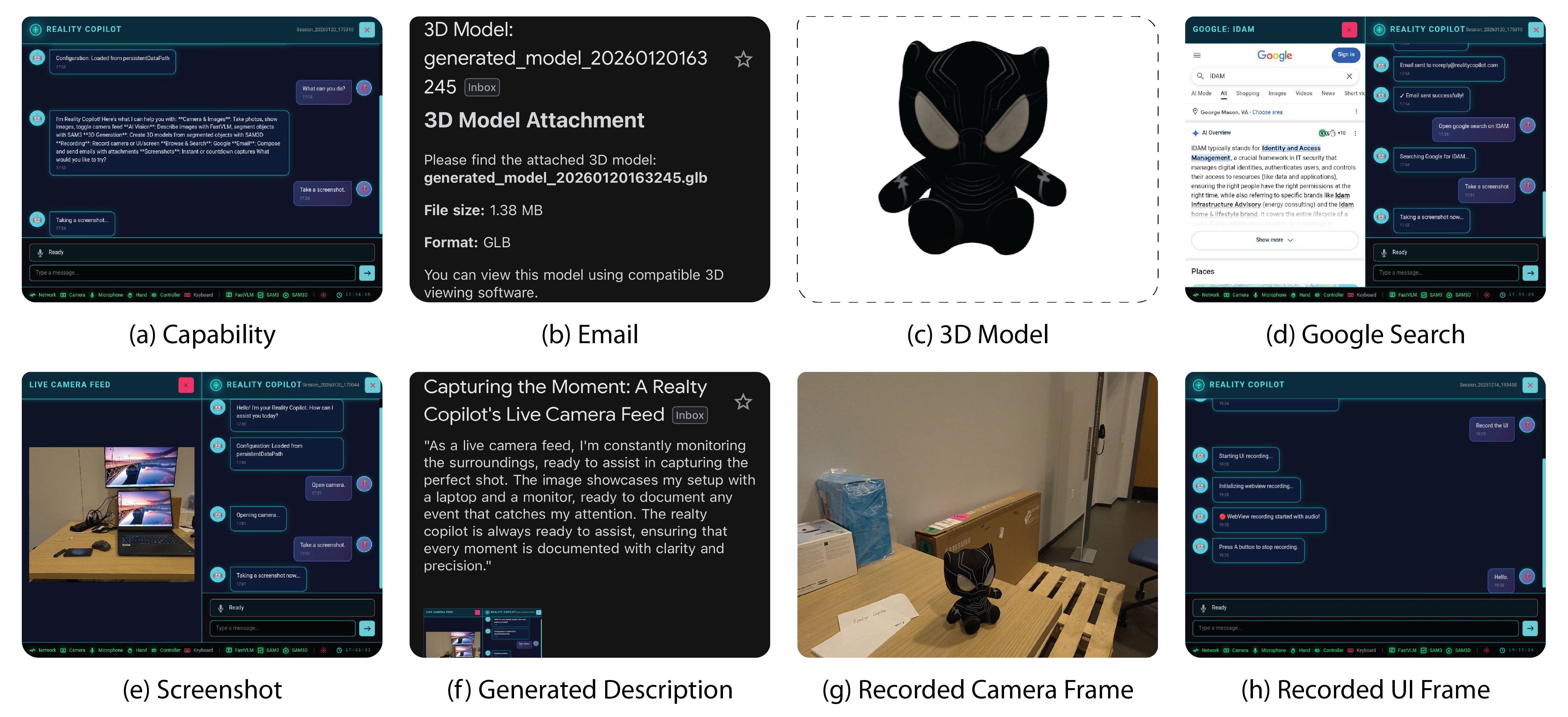}
  \caption{Overview of Reality Copilot functionalities. (a) The main entry window, which includes the title, chat box, voice input, and status bar, along with a response for the system capability question. This window dynamically adapts based on user interactions, as illustrated in (d) and (e). (b) An email containing a 3D model generated using the SAM3D model. (c) The corresponding 3D model downloaded from the email attachment. (d) An example of integrated Google Search results. (e) The screenshot functionality while the camera is open. (f) An email with a generated title and content using the FastVLM model. (g) A frame from the recorded egocentric headset camera video. (h) A frame from the recorded user interface video.}
  \label{fig:teaser}
\end{teaserfigure}


\maketitle

\section{Introduction}



Large Language Models (LLMs) have gained widespread adoption following the release of ChatGPT~\cite{wu2023brief}. Text-based interactions have enabled a wide range of applications, including Wikipedia-style information retrieval, issue diagnosis and resolution, and programming problem solving. Building on this foundation, Large Multimodal Models (LMMs) extend the interaction paradigm by incorporating mixed inputs such as text, audio, images, and video~\cite{yin2024survey}, thereby enriching human-AI interaction. For instance, Google Nano Banana can generate images from text prompts, while OpenAI Sora can convert text and images into video. Despite these advances, interaction with LLMs and LMMs remains confined to traditional interfaces such as web browsers or standalone applications on PCs and smartphones.



Smart wearables, such as mixed reality headsets, are increasingly being adopted as productivity tools. For example, Mac Virtual Display~\footnote{\url{https://support.apple.com/en-us/118521}} enables the Apple Vision Pro to function as a private, portable 5K display for MacBook devices. Similarly, Meta Remote Desktop~\footnote{\url{https://www.meta.com/help/quest/1370025034331518/}} allows users to extend their computer setup with multiple large virtual displays using the Quest 3 headset. Beyond remote streaming and display capabilities, mixed reality headsets also support productivity applications such as Microsoft Office. However, these applications often replicate desktop interfaces without adapting to the unique affordances and constraints of head-mounted displays. For instance, hand- or controller-based interactions in mixed reality can be significantly more fatiguing than traditional mouse and keyboard input on a PC.


In response to these challenges, we introduce \textit{Reality Copilot}, a voice-first human-AI collaboration assistant for mixed reality, powered by Large Multimodal Models. Reality Copilot leverages voice interaction as the primary user interface, enabling hands-free operation in immersive environments. It employs a hybrid architecture that integrates both commercial large-scale models (e.g., OpenAI ChatGPT and Google Gemini) and open-source models (e.g., SAM3~\cite{carion2025sam3segmentconcepts}, SAM3D~\cite{sam3dteam2025sam3d3dfyimages}, and FastVLM~\cite{vasu2025fastvlm}). This design allows for privacy-preserving processing, where sensitive data, such as images, is handled locally via self-hosted servers.

To bridge the interaction gap between traditional and immersive interfaces, Reality Copilot offers familiar functionalities such as screenshot capture and screen recording. Leveraging raw camera access and contextual awareness from the headset hardware, it can perceive real-world surroundings to inform intelligent assistance. Furthermore, with hardware-accelerated video encoding, Reality Copilot supports real-time recording of both the user interface and the external environment, along with dual-track audio capture from the microphone and speaker.



The contributions of this work are as follows:

\begin{itemize}
    \item Proposing \textit{Reality Copilot}, a voice-first human-AI collaboration framework for mixed reality that integrates both commercial and open-source Large Multimodal Models.
    \item Presenting the detailed implementation of a fully functional AI assistant tailored for mixed reality, leveraging hardware capabilities such as raw camera access and hardware-accelerated video encoding.
    \item Demonstrating a range of use cases and application scenarios that highlight the versatility and potential of Reality Copilot in everyday productivity and immersive computing contexts.
\end{itemize}

\section{Related Work}


\subsection{Human-AI Collaboration in Mixed Reality}

A growing body of work has explored how AI systems can augment human performance in mixed reality environments. For instance, Dang et al.~\cite{dang2025human} investigated human-AI collaboration in educational contexts, focusing on interactions between learners and embodied generative AI agents. Liao et al.~\cite{liao2022realitytalk} introduced RealityTalk, a real-time speech-driven system that empowers users to deliver live augmented presentations through natural voice interaction. In the domain of creative expression, Wang and Martin~\cite{wang2025ai} proposed a co-creative musical interface that visualizes AI-generated musical actions, highlighting collaborative dynamics between humans and AI in mixed reality. Furthermore, Tang et al.~\cite{tang2023real} developed a digital twin system that integrates object recognition and mixed reality for real-time data streaming, demonstrating the utility of AI-augmented spatial computing. In the medical domain, Moglia et al.~\cite{moglia2023mixed} presented HoloKnee, a system that combines mixed reality and AI for multimodal data visualization and surgical planning in knee osteotomy.

For a broader overview of AI-enabled applications in mixed reality, please refer to the comprehensive survey by Hirzle et al.~\cite{hirzle2023xr}.

\subsection{Large Multimodal Model Applications}

In parallel, rapid advancements in Large Multimodal Models (LMMs), more specifically, Multimodal Large Language Models (MLLMs), have led to a surge in applications across a wide range of domains. These models enable rich interactions across text, vision, audio, and video inputs, opening new opportunities for human-AI collaboration. Recent studies have demonstrated the applicability of LMMs in diverse fields such as healthcare~\cite{alsaad2024multimodal, mesko2023impact}, well-being~\cite{yu2025floatmind}, construction~\cite{erfani2026applications}, intelligent agents~\cite{xie2024large}, autonomous driving~\cite{cui2024survey, li2025applications}, medicine~\cite{qiu2024application}, object detection~\cite{li2025lmm}, video understanding and editing~\cite{team2025vidi}, pixel-level reasoning~\cite{ren2024pixellm}, and document understanding~\cite{liu2024textmonkey}.

Specifically, Liu et al.~\cite{liu20233dall} introduced 3DALL-E, a text-to-image plugin for CAD environments that integrates DALL-E, GPT-3, and CLIP, enabling designers to generate visual concepts based on contextual prompts. Their findings suggest that such tools enhance collaboration and stimulate design exploration. Similarly, Wang et al.~\cite{wang2025aideation} proposed AIdeation, a system that leverages multiple generative models to support early-phase concept ideation. Their user study showed that AIdeation significantly boosted creativity, idea diversity, and efficiency compared to traditional workflows.

For further insights into the growing landscape of LMM applications, please refer to recent surveys and reviews~\cite{wu2023multimodal, liang2024survey, tu2024overview}.

\section{Overview}

\begin{figure}[!t]
    \centering
    \includegraphics[width=\linewidth]{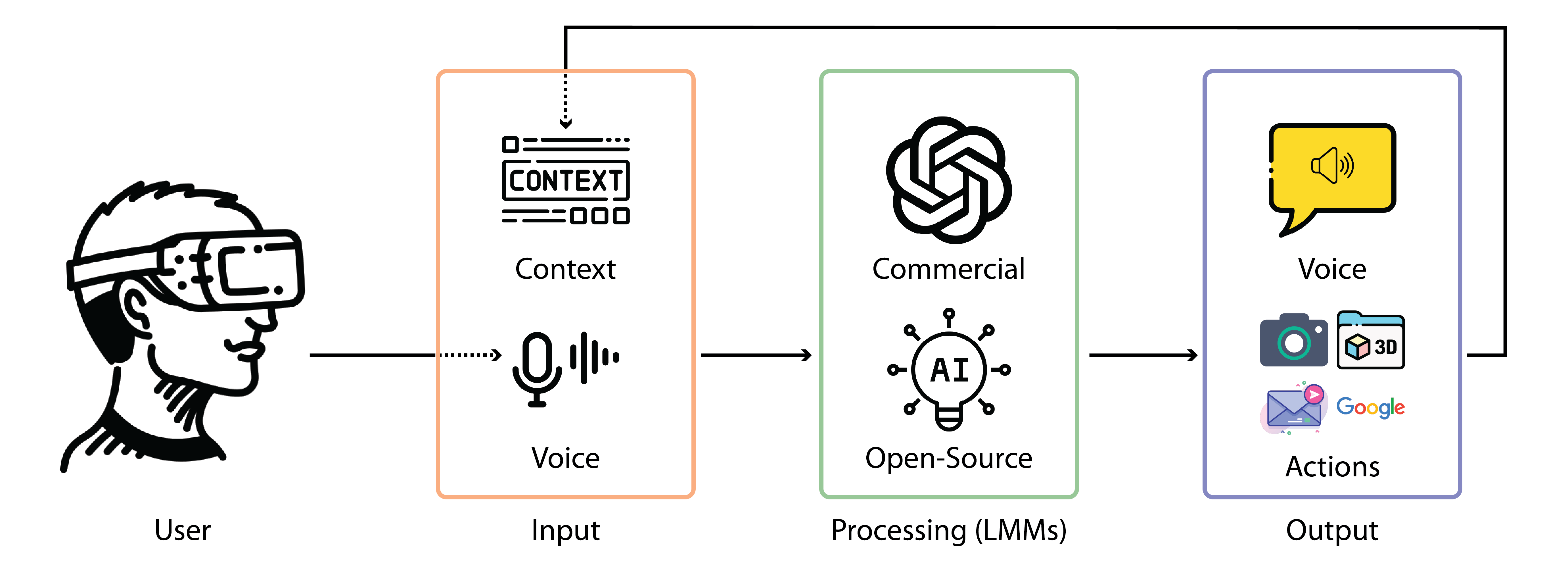}
    \caption{Workflow of Reality Copilot. When a user wears a mixed reality headset and launches Reality Copilot, they can interact using natural voice. The voice input, along with contextual information (e.g., user interface state and service availability), is sent to the LMMs. Reality Copilot integrates two types of LMMs: commercial and open-source. Voice inputs are processed by commercial LMMs, while image and 3D model processing is handled by open-source LMMs. The output consists of both voice responses and system actions, which are also used to update the internal context.}
    \vspace{-15mm}
    \label{fig:overview}
\end{figure}

The system workflow is illustrated in Fig.~\ref{fig:overview}. When a user wears a mixed reality headset (e.g., Meta Quest 3), they can interact with \textit{Reality Copilot} using natural voice. The captured voice input, along with contextual information, such as network status, user interface state, and service availability, is sent to an LMM, such as OpenAI ChatGPT or Google Gemini. The LMM analyzes the user’s intent and generates a structured response guided by a customized system prompt. If the model fails to interpret the intent, it proactively prompts the user for clarification.

The system response consists primarily of two types: \textit{voice} and \textit{action}. Voice responses are played through the headset’s internal speaker, while actions are directly executed by the system. For example, if the action is to \textit{open the camera}, the headset’s camera is activated. If the action is to \textit{describe the image}, the FastVLM service is invoked to generate a visual description. If the action is to \textit{generate a 3D model}, the SAM3D service is triggered to synthesize a corresponding 3D object.

Additionally, the output is used to update the internal context, enabling context-aware interaction. We implement a stack-based context processing mechanism to manage the flow of user interactions over time. Although Reality Copilot is designed as a voice-first system, text input is also supported.

\begin{figure}[t]
    \centering
    \begin{subfigure}[t]{0.49\linewidth}
        \centering
        \includegraphics[width=\textwidth]{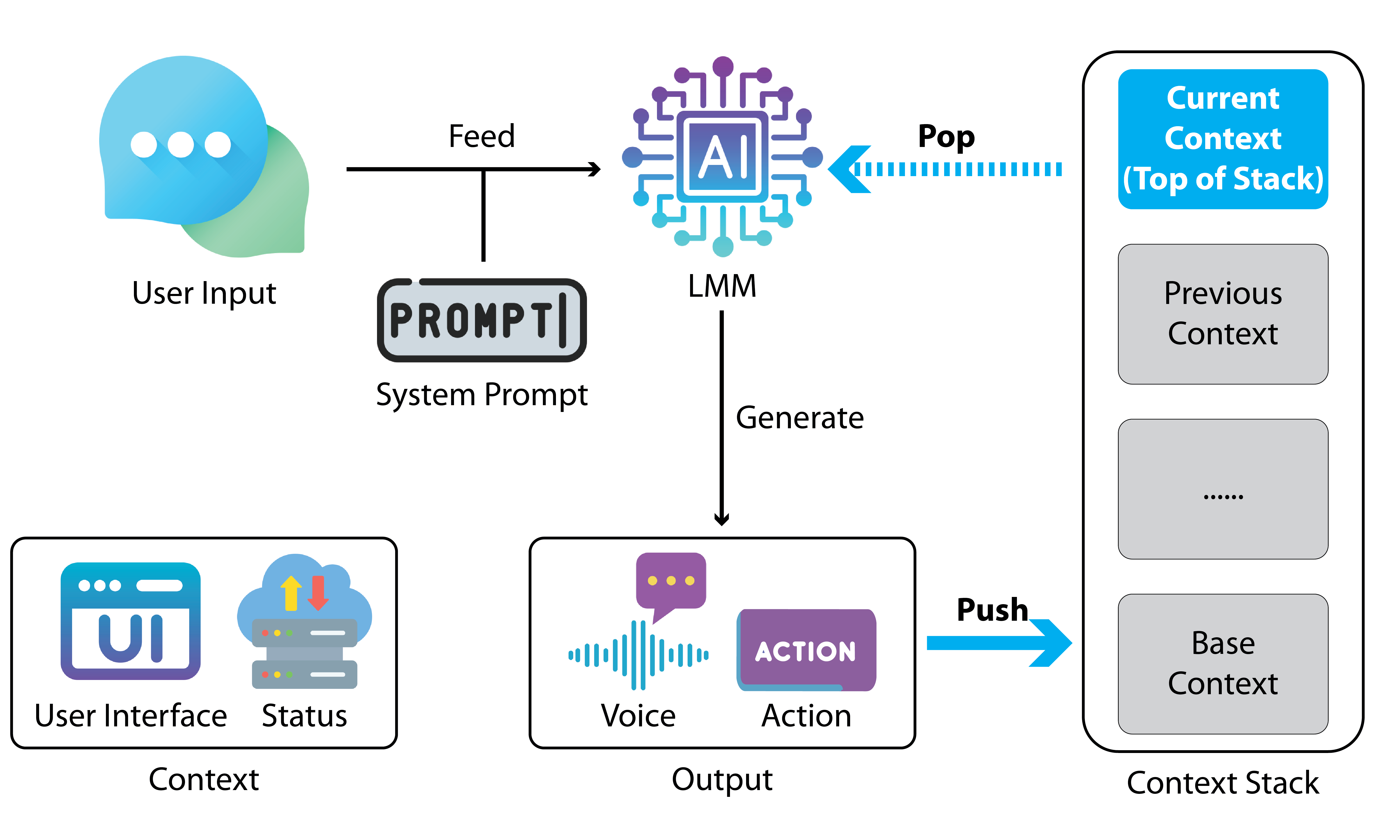}
        \caption{Stack-based context processing. The context includes the user interface state and system status, such as network connectivity and backend service availability. When a user input is about to be sent to the LMM, the system checks its relevance to the current context. If relevant, the context is popped, together with the user input and system prompt, processed by the LMM using a customized prompt. The LMM then outputs a voice response and corresponding actions. After the voice is played and the actions are executed, the context stack is updated accordingly.}
        \label{fig:stack}
    \end{subfigure}
    \hfill
    \begin{subfigure}[t]{0.49\linewidth}
        \centering
        \includegraphics[width=\textwidth]{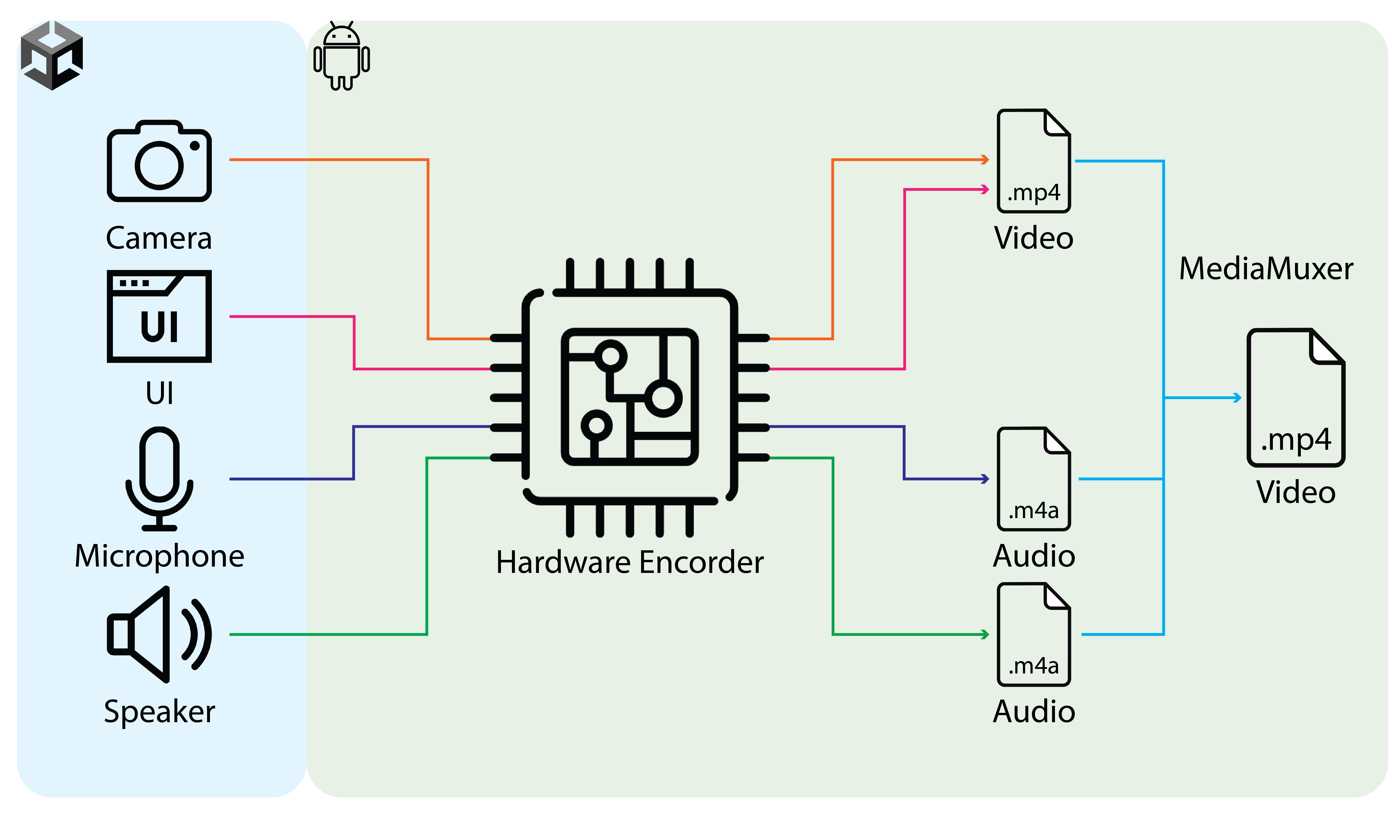}
        \caption{Hardware-accelerated recording pipeline. When camera or UI recording is triggered from the Unity side, it activates the hardware encoder on the Android side. The video stream, along with audio from the microphone and speaker, is recorded separately. Video is encoded using the H.264 codec and saved as an MP4 file, while audio from the microphone and speaker is encoded using AAC and saved as separate M4A files. Once recording is stopped, all three streams are muxed into a single MP4 file using \texttt{MediaMuxer}.}
        \label{fig:recording}
    \end{subfigure}
    \vspace{-2mm}
    \caption{System internals of Reality Copilot: (a) Stack-based context processing; (b) Hardware-accelerated recording pipeline.}
    \vspace{-5mm}
    \label{fig:stack_recording}
\end{figure}

The stack-based context processing workflow is depicted in Fig.~\ref{fig:stack_recording} (a). The context currently encompasses elements such as the user interface state (e.g., what is being displayed) and system status (e.g., availability of open-source LMM services). When a new user input is about to be sent to the LMM, the most recent context is popped from the context stack. The LMM then generates a structured output, comprising a voice response and a set of actions based on both the user input and the retrieved context. Once the output is processed, the updated context is pushed back onto the stack.

In addition, users can explicitly trigger a stack-pop operation via voice command. For example, if a photo is currently being displayed and the user says, ``email the image to me,'' the system uses the current user interface context (i.e., \textit{photo}) and status context (e.g., whether the FastVLM service is available). If FastVLM is available, it is used to automatically generate the email title and content. Otherwise, the system falls back to using the image filename. The email window and content are then pushed onto the context stack.

Notably, if the user input is determined to be irrelevant to the current context, a stack-pop operation won't be performed. This design helps maintain coherence and avoids unnecessary context switching during interaction.



\section{Implementation}

\textit{Reality Copilot} is implemented on Windows 11 using Unity 6000.2.6f2. Android Studio 2025.2 is employed to develop an Android plugin for video recording. Reality Copilot is deployed as a standalone application on Meta Quest 3 headsets. The open-source backend services~\footnote{\url{https://github.com/luffy-yu/RealityCopilot-Backend}} are hosted on an Ubuntu 24.04 desktop equipped with an NVIDIA GTX 5090 GPU.

\ssection{Main User Interface.}  
The main user interface is built using Vuplex 3D WebView~\footnote{\url{https://developer.vuplex.com/webview/overview}}, which enables integration between Unity and HTML in a hybrid framework. To support voice-first, hands-free interaction, we integrate TEN VAD~\cite{TENVAD}, a real-time voice activity detection system that eliminates the need for manual voice input toggling.

\ssection{Large Multimodal Models Integration.}  
For LMM integration, we adopt EmBARDiment~\cite{bovo2025embardiment}, a toolkit designed for AI agents in Android XR environments, which uses Google Gemini as the backend. We extend the framework to also support OpenAI APIs. Additionally, we integrate open-source models, FastVLM~\cite{vasu2025fastvlm}, SAM3~\cite{carion2025sam3segmentconcepts}, and SAM3D~\cite{sam3dteam2025sam3d3dfyimages}, for image understanding and 3D model generation on a self-hosted server.

\ssection{Headset Camera Access.}  
Camera access is provided through the official Passthrough Camera Access (PCA) component from the Mixed Reality Utility Kit (MRUK), enabling real-time scene capture within the headset.

\ssection{Hardware-Accelerated Video Recording.}  
We extend an Android plugin~\cite{zhao2025xrobotoolkit} to support dual-track audio recording from both the microphone and speaker. The audio is encoded using AAC, while the video stream is encoded using H.264. These are muxed into a single video file with separate audio tracks and stored in the headset’s internal storage. The recording pipeline is illustrated in Figure~\ref{fig:stack_recording} (b).

\ssection{Email Functionality.}  
Email functionality is implemented using the Gmail API, allowing Reality Copilot to automatically generate and send context-aware emails.

\vspace{-2mm}
\section{Applications}








The voice-first feature of \textit{Reality Copilot} unlocks a wide range of possibilities by enabling hands-free interaction. 

\vspace{-2mm}
\subsection{Real-Time Assistant}

The integration of Large Multimodal Models allows Reality Copilot to provide real-time assistance in various scenarios. For instance, a student learning electronics can converse directly with Reality Copilot to receive step-by-step guidance on tasks such as circuit specifications, soldering procedures, and wire connection checks (Fig.~\ref{fig:applications} (a)). This hands-free support enhances learning efficiency and safety during hands-on activities.

\vspace{-2mm}
\subsection{Egocentric Video Creation}

\begin{figure}[t]
    \centering
    \includegraphics[width=\linewidth]{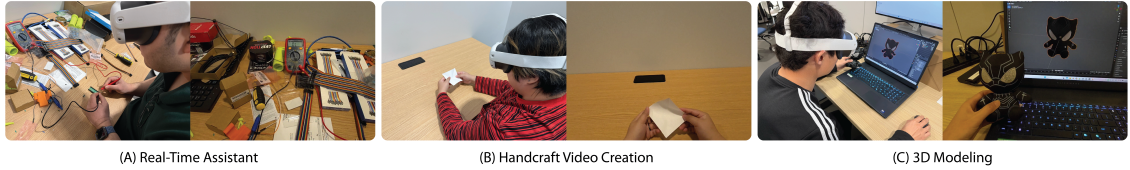}
    \vspace{-7mm}
    \caption{Application samples of Reality Copilot. In each subfigure, the left-side image shows a third-person view captured with an iPhone, while the right-side image presents the corresponding egocentric (user) view. (A) Reality Copilot, powered by LMMs, assists students in learning electronics by providing real-time, voice-driven guidance. (B) Reality Copilot supports egocentric video creation with real-time narration, aiding content creators in documenting handcrafting processes. (C) Reality Copilot enhances the 3D modeling workflow by enabling designers to generate realistic 3D models.}
    \vspace{-5mm}
    \label{fig:applications}
\end{figure}

Reality Copilot supports egocentric video recording with dual audio tracks, making it a powerful tool for content creation. For example, a craftsperson can use the system to document their process while narrating their actions, with the voiceover automatically recorded alongside the video (Fig.~\ref{fig:applications} (b)). Additionally, Reality Copilot can be integrated into e-commerce live streaming scenarios~\cite{liao2022realitytalk}, enhancing remote product showcasing and audience engagement.

\vspace{-2mm}
\subsection{3D Modeling Workflow Integration}

With built-in support for 3D model generation via SAM3D~\cite{sam3dteam2025sam3d3dfyimages}, Reality Copilot can assist designers in rapidly creating realistic 3D models. The generated GLB files can be automatically emailed to the user, enabling seamless integration into their design workflow (Fig.~\ref{fig:applications} (c)). These models can be directly imported into applications such as Blender or Fusion~360, including design tools such as the text-to-image plugin for 3D CAD~\cite{liu20233dall}.

\section{Conclusion}


We present \textit{Reality Copilot}, a voice-first human-AI collaboration prototype in mixed reality, powered by Large Multimodal Models. Reality Copilot adopts a hybrid architecture that integrates both commercial solutions and open-source models. To preserve user privacy, image and 3D model processing is performed locally on self-hosted servers. Reality Copilot enables real-time assistance for a range of tasks, including information retrieval, context-aware interaction, 3D model generation, online search, and email composition. We detail the system's implementation and demonstrate its potential through three illustrative application scenarios. To the best of our knowledge, Reality Copilot represents the first exploration of voice-first human-AI collaboration in mixed reality powered by LMMs. We believe this work offers valuable insights and opens new directions for future work in immersive multimodal human-AI interaction.

\begin{acks}
This project was supported by the National Science Foundation (award numbers: 2430673, 2418236, and 1942531).
\end{acks}

\bibliographystyle{ACM-Reference-Format}
\bibliography{main}




\end{document}